\documentclass[draftclsnofoot,onecolumn]{IEEEtran}

\usepackage{epsfig,rotating,setspace,latexsym,amsfonts}
\usepackage{algorithm,algorithmic,graphicx,epsf,authblk,epstopdf,url,color,multirow}
\usepackage{tabularx}
\usepackage{amsthm,amsmath,amssymb,cite}
\usepackage{eucal}
\usepackage{mathtools}
\usepackage{enumerate}
\usepackage{xspace}
\usepackage{bm}
\usepackage{lastpage}
\usepackage{caption}
\usepackage{xcolor}
\allowdisplaybreaks
\usepackage{enumitem}

\newtheorem{theorem}{Theorem}
\newtheorem{lemma}{Lemma}

\newtheorem{remark}{Remark}

\newtheorem{proposition}{Proposition}

\DeclareMathAlphabet{\mathcal}{OMS}{cmsy}{m}{n}

\usepackage{graphicx}
\usepackage[font=small]{caption}
\usepackage{subcaption}

\begin{document}

\title{On the Upload versus Download Cost for Secure and Private Matrix Multiplication}

\author{
Wei-Ting~Chang \qquad Ravi~Tandon\\ 
Department of Electrical and Computer Engineering\\
University of Arizona, Tucson, AZ, USA\\
E-mail: \{\textit{wchang, tandonr}\}@email.arizona.edu
}


\maketitle

\begin{abstract}
In this paper, we study the problem of secure and private distributed matrix multiplication. Specifically, we focus on a scenario where a user wants to compute the product of a confidential matrix $A$, with a matrix $B_\theta$, where $\theta\in\{1,\dots,M\}$. The set of candidate matrices $\{B_1,\dots,B_M\}$ are public, and available at all the $N$ servers. The goal of the user is to distributedly compute $AB_{\theta}$, such that $(a)$ no information is leaked about the matrix $A$ to any server; and $(b)$  the index $\theta$ is kept private from each server. Our goal is to understand the fundamental tradeoff between the upload vs download cost for this problem. 
Our main contribution is to show that the lower convex hull of following (upload, download) pairs: $(U,D)=(N/(K-1),(K/(K-1))(1+(K/N)+\dots+(K/N)^{M-1}))$ for $K=2,\dots,N$ is achievable. The scheme improves upon state-of-the-art existing schemes for this problem, and leverages ideas from secret sharing and coded private information retrieval.

\end{abstract}

\section{Introduction}
\footnote{This work was supported by NSF Grants CAREER 1651492, and CNS 1715947.}
Distributed computing aims at speeding up computationally intensive operations by dividing and outsourcing the computation to multiple servers. Distributed processing often comes with additional communication overhead, when compared to centralized processing. Furthermore, when processing sensitive information in a distributed manner, security and privacy are also critical concerns.  Hence,  simultaneously utilizing distributed processing, while satisfying security/privacy constraints are critical and of great interest.

A well known issue in distributed systems is the \textit{straggler effect}, where slower machines in the system slow down the entire computation. Recently, error control codes have been successfully adopted to carefully add computational redundancy in order to tackle the stragglers problem. For instance, entangled polynomial codes \cite{PolyCode2018} and PolyDot codes \cite{PolyDot} have been proposed for the stragglers problem, within the context of distributed matrix multiplication.

\begin{figure}[t]
\centering
	\includegraphics[width=0.65\linewidth]{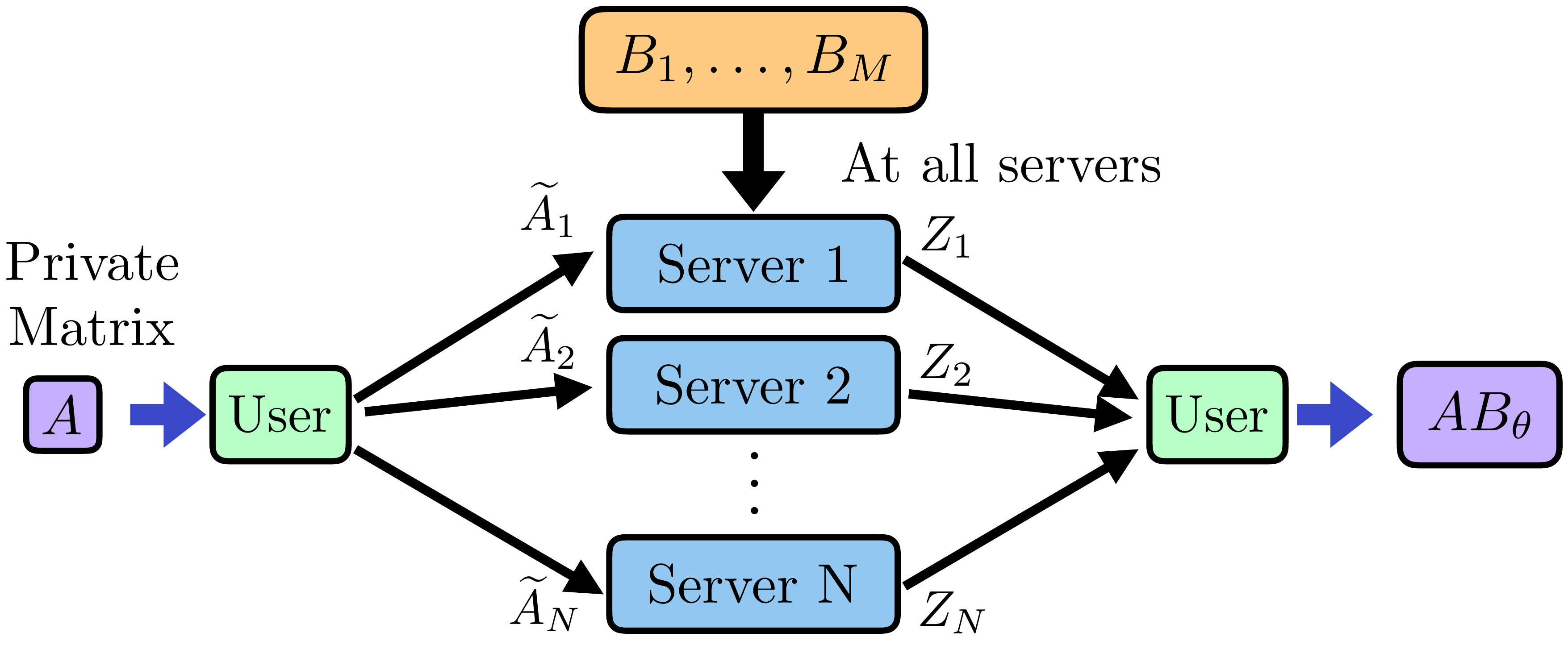}
	\caption{System model for secure and private matrix multiplication: multiply a confidential matrix $A$ with $B_\theta$, while keeping the index $\theta$ private from all $N$ servers.
    \label{fig:model}}
	\vspace{-20pt}
\end{figure}

Although mitigating stragglers is important, it is crucial to ensure that the data, if confidential, is not leaked to unauthorized servers. Recently, another line of research has focused on the security aspects for the distributed matrix multiplication problem. Secure matrix-vector multiplication was considered in \cite{Salim2017}. In \cite{MaddahAli-ISIT-2018}, the authors considered a variation of the problem where the matrices are kept secure against the servers as well as the user. The authors in \cite{Lagrange2018} proposed the idea of Lagrange coded computing that simultaneously deals with straggler, colluding servers and malicious servers. Later, \cite{ChangTandon,YangLee} studied the problem of secure distributed matrix multiplication. Both works proposed  schemes based on polynomial codes, where the authors of \cite{YangLee} studied the non-colluding version of the problem and the authors of \cite{ChangTandon} studied the colluding version (also see recent improvements and variants in \cite{Kakar, GASP, SimeoneSMM, ChangTandonFFT}).  


Another issue that needs to be addressed is privacy of the user. Consider a scenario where in addition to the confidential data, the queries sent for computation are also sensitive, i.e., the queries can leak private information about the user. 
The authors of \cite{KimLeePSMM} considered both security and privacy for the problem of distributed matrix multiplication. There are several other works that have focused primarily on the privacy aspect. In particular, starting from the work of \cite{SunJafar}, private information retrieval (PIR) has been studied extensively within an information theoretic framework (such as PIR from coded databses \cite{Banawan} and private computation \cite{PC2017Jafar,PFR2017MA,PC2018Karpuk,PFR2018MDS}).

\textit{Main Contributions:} In this paper, we focus on the model where a user wants to compute the product of a confidential matrix $A$, with a matrix $B_\theta$, where $\theta\in\{1,\dots,M\}$. The set of candidate matrices $\{B_1,\dots,B_M\}$ are public, and available at all the $N$ servers. The goal of the user is to distributedly compute $AB_{\theta}$, such that $(a)$ no information is leaked about the matrix $A$ to any server; and $(b)$  the index $\theta$ is kept private from each server.  Our main contribution is to show that the lower convex hull of following (upload, download) pairs: $(U,D)=(N/(K-1),(K/(K-1))(1+(K/N)+\dots+(K/N)^{M-1}))$ for $K=2,\dots,N$ is achievable. Our scheme also improves upon the scheme of \cite{KimLeePSMM}.

\section{Problem Formulation \label{sec:systemModel}}
We consider a distributed system with one user and $N$ non-colluding servers. The user has a confidential matrix $A\in\mathbb{F}^{d_1\times d_2}$, and all servers have access to $M$ matrices $B_m\in \mathbb{F}^{d_2\times d_3},~m=1,\dots,M$, for some integers $d_1,d_2$ and $d_3$, and a sufficiently large field $\mathbb{F}$. We assume that the matrices $A, B_1, \cdots, B_M$ are all independent of each other, and each with i.i.d. entries from $\mathbb{F}$. The goal of the user is to compute the product of a confidential matrix $A$ with $B_\theta$, where the index $\theta$ is private. Each server is connected to the user through a separate link. We assume that all servers are honest but curious (i.e., all servers correctly follow the protocols, however, they are interested in learning about $A$ and $\theta$). To ensure security, the user securely encodes $A$ using encoding functions $\bm{f}=(\mathit{f}_1,\dots,\mathit{f}_N)$, where $\mathit{f}_n$ is the individual encoding function for server $n$. We denote the encoded version of $A$ that is sent to server $n$ by $\widetilde{A}_n$, i.e., $\widetilde{A}_n=\mathit{f}_n(A)$. Along with $\widetilde{A}_n$, the user also sends a query $Q_n^{(\theta)}$ to server $n$. 
Without any prior knowledge of the stored data, the queries sent by the user are independent of all the $B_m$'s, i.e., $I(Q_1^{(\theta)},\dots,Q_N^{(\theta)};B_1,\dots,B_M)=0$. Each server $n$ uses a computing function $\mathcal{Z}_n^{(\theta)}:\{B_1,\dots,B_M\}\times \widetilde{A}_n\times Q_n^{(\theta)}\rightarrow Z_n^{(\theta)}$ for the assigned task. Once servers finish the requested computations, servers return their answers $Z_n^{(\theta)},~n=1,\dots,N,$ to the user. Next, to preserve the privacy of the user, the strategy the user used to download the desired result should not reveal $\theta$. That is, for any $\theta$ and $\theta'$, the query and the answer from server $n$ should be statistically identical. Upon receiving all the $Z_n^{(\theta)}$'s, the user decodes the desired result by using decoding function, $AB_{\theta}= g(Z_1^{(\theta)} , Q_{1}^{(\theta)},\dots, Z_N^{(\theta)},Q_{N}^{(\theta)})$. A scheme is secure, private and reliable if it satisfies the following constraints:

\noindent \textbf{Security Constraint:}
\begin{align}
I(A;\widetilde{A}_n, Q_n^{(\theta)},B_1,\dots,B_M)=0,~n=1,\dots,N.
\end{align}

\noindent \textbf{Privacy Constraint:}
\begin{align}
\hspace{-2mm}(\widetilde{A}_n,Q_n^{(\theta)},Z_n^{(\theta)},B_{[1:M]}) \sim (\widetilde{A}_n,Q_n^{(\theta')},Z_n^{(\theta')},B_{[1:M]}),\forall n
\end{align}

\noindent \textbf{Decodability Constraint:}
\begin{align}
H\left(AB_\theta|Z_1^{(\theta)},Q_{1}^{(\theta)}, \cdots, Z_N^{(\theta)}, Q_{N}^{(\theta)}\right)=0.
\end{align}

The performance of a scheme is determined by the normalized upload cost $(U)$ and download cost $(D)$, defined as:
\begin{align}
U=\frac{\sum\limits_{n=1}^{N}H(\widetilde{A}_n)}{H(AB_\theta)},\text{ and } D=\frac{\sum\limits_{n=1}^{N}H(Z_n^{(\theta)})}{H(AB_\theta)}.
\end{align}
Our aim is to understand the tradeoff between the upload and download costs, i.e., the set of all feasible $(U,D)$ pairs. The optimal download cost for a fixed upload cost is defined as:
\begin{align}
D^*(U) \triangleq \min\{D:(U,D)\text{ is feasible}\}.
\end{align}
The following Lemma shows that the optimal download cost is a convex function of the upload cost.

\begin{lemma}
The optimal download cost $D^*(U)$ is a convex function of the upload cost $U$.
\label{lemma:optimalD}
\end{lemma}
\noindent \textbf{Proof:} To prove this Lemma, we show that for any two upload costs $U_1,U_2$ and their corresponding optimal download costs $D^*(U_1),D^*(U_2)$, there is a scheme with upload cost $\bar{U}=\alpha U_1+(1-\alpha)U_2$ that achieves a download cost of $\bar{D}=\alpha D^*(U_1)+(1-\alpha)D^*(U_2), 0\leq\alpha\leq 1$. This can be shown by memory sharing argument where we partition $A$ into two parts, $A^{(\alpha)}\in \mathbb{F}^{\alpha d_1\times d_2}$ and $A^{(1-\alpha)}\in \mathbb{F}^{(1-\alpha) d_1\times d_2}$, respectively. $A^{(\alpha)}$ is sent to servers securely using scheme $1$ with upload cost $U_1$. The answers are returned to the user using the optimal scheme that corresponds to $U_1$, hence, achieves the download cost of $D^*(U_1)$. Similarly, $A^{(1-\alpha)}$ is sent to servers and the answers are returned to the user using scheme $2$ that achieves $(U_2,D^*(U_2))$. The desired results obtained from both parts are of size $\alpha d_1\times d_3$ and $(1-\alpha) d_1\times d_3$. Since these are per-bit costs,  the total upload and download costs are
\begin{align}
\hspace{-3mm}U_{\text{total}}&=\alpha U_1 d_1d_2 + (1-\alpha) U_2 d_1d_2=\bar{U}d_1d_2,\\
\hspace{-3mm}D_{\text{total}}&=\alpha D^*(U_1) d_1d_3 + (1-\alpha)D^*(U_2) d_1d_3=\bar{D}d_1d_3.
\end{align}
Clearly, the optimal $D^*(\bar{U})$ is upper bounded by $\bar{D}(\bar{U})$ completing the proof of Lemma \ref{lemma:optimalD}.

In the next proposition, we show that even without any privacy constraints, the minimum upload cost is $N/(N-1)$. 
\begin{proposition}
The minimum value of upload cost is $U_{\text{min}}=N/(N-1)$.
\label{pro:1}
\end{proposition}
\noindent \textbf{Proof:} We start with the following sequence of inequalities:
\begin{align}
\sum_{n=1}^{N}H(\widetilde{A}_n) &\geq H(\widetilde{A}_1,\dots,\widetilde{A}_N)\nonumber\\
&\stackrel{(a)}{=} H(\widetilde{A}_1,\dots,\widetilde{A}_N|B_1,\dots,B_M)\nonumber\\
&\stackrel{(b)}{=} H(AB_\theta,\widetilde{A}_1,\dots,\widetilde{A}_N|B_1,\dots,B_M)\nonumber\\ 
&= H(AB_\theta|B_1,\dots,B_M) + H(\widetilde{A}_1,\dots,\widetilde{A}_N|AB_\theta,B_1,\dots,B_M)\nonumber\\
&\stackrel{(c)}{\geq} H(AB_\theta) + H(\widetilde{A}_n|AB_\theta,B_1,\dots,B_M)\nonumber\\
&\stackrel{(d)}{=} H(AB_\theta) + H(\widetilde{A}_n|B_1,\dots,B_M)\nonumber\\
&\stackrel{(e)}{=} H(AB_\theta) + H(\widetilde{A}_n)\label{uploadone}
\end{align}
where $(a)$ follows from the fact that $\widetilde{A}_n$'s are independent of $B_m$'s for all $m,n$; $(b)$ follows from decodability constraint; $(c)$ is due to the fact that $H(AB_\theta|B_1,\dots,B_M)=H(AB_\theta)$ by security constraint; $(d)$ also follows from security constraint and the fact that $A,B_1,\dots,B_M$ are all independent of each other so that $AB_\theta$ can be removed from the conditioning; $(e)$ follows due to independence of $\widetilde{A}_n$ and $\{B_1, \ldots, B_M\}$. Summing up (\ref{uploadone}) for all $n$, we have
$N(\sum_{n=1}^{N}H(\widetilde{A}_n)) \geq NH(AB_\theta) + \sum_{n=1}^{N}H(\widetilde{A}_n)$. By rearranging, we obtain 
\begin{align}
U=\frac{\sum_{n=1}^{N}H(\widetilde{A}_n)}{H(AB_\theta)} \geq \frac{N}{N-1}, 
\end{align}
completing the proof of Proposition \ref{pro:1}.

\newpage 
\section{Main Results and Discussion}
\begin{theorem}
For the secure and private matrix multiplication problem with $N$ non-colluding servers, where each server has access to $M$ public matrices, the lower convex hull of
\begin{align}
(U,D) = \left(\frac{N}{K-1}, \frac{K}{K-1}\left(1+\frac{K}{N}+\dots+\left(\frac{K}{N}\right)^{M-1}\right)\right)\nonumber
\end{align} 
for $K=2,\dots,N$ is achievable.
\label{thm:updown}
\end{theorem}

\begin{remark}
In Theorem \ref{thm:updown}, by varying the parameter $K$, we can trade off upload cost for download cost. We note that for minimum upload cost, i.e., when $K=N$ and $U=N/(N-1)$, and $M=1$ (i.e., no privacy constraint), the download cost of $D=N/(N-1)$ is information-theoretically optimal as shown in our previous work \cite{ChangTandon}.
\end{remark}

\begin{figure}[t]
\centering
	\includegraphics[width=0.55\linewidth]{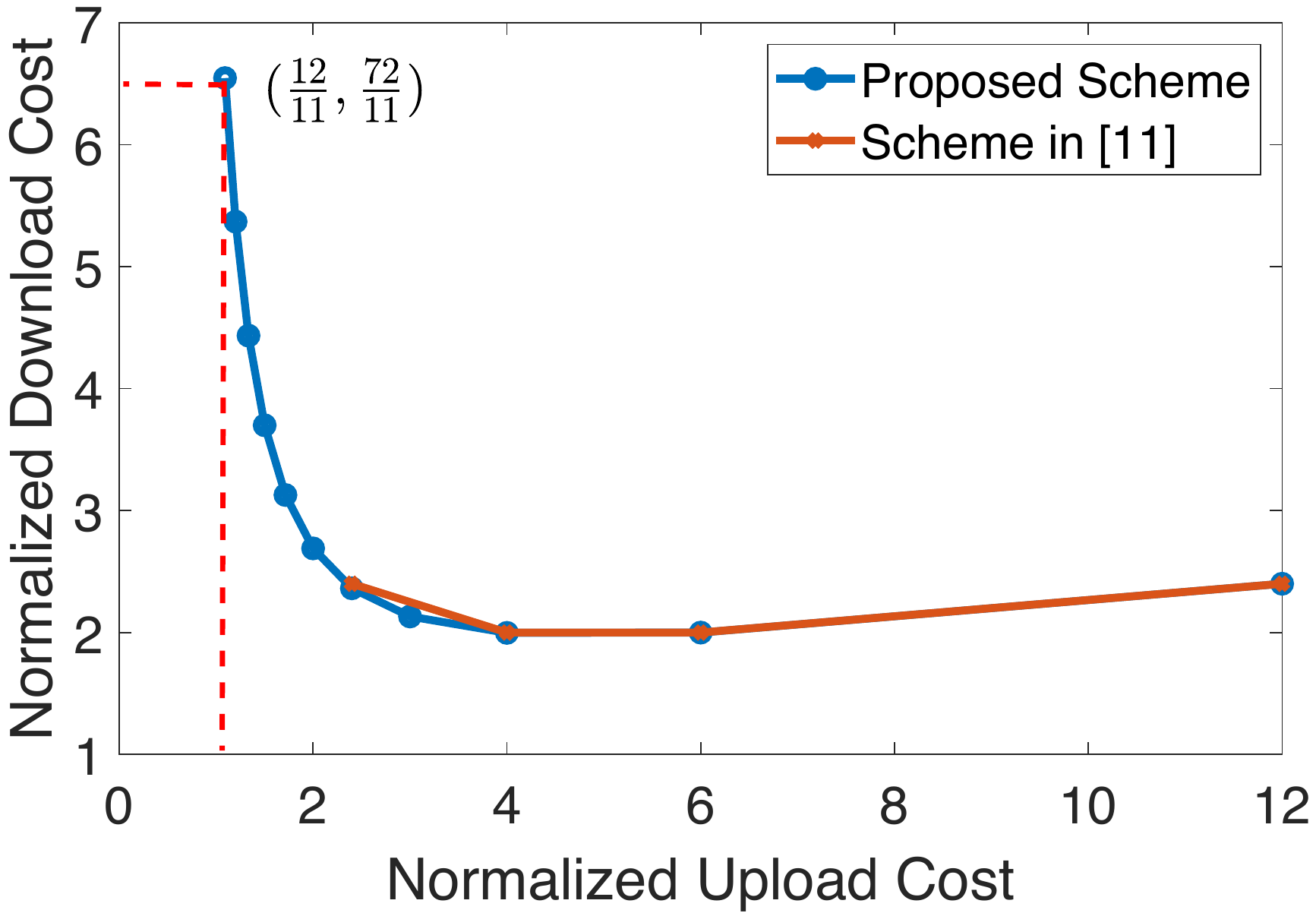}
	\caption{Comparison between our proposed scheme and the scheme in \cite{KimLeePSMM}, with $N=12,M=6$ and $K=2,3,\dots,12$.
    \label{fig:compare}}
	\vspace{-20pt}
\end{figure}

\begin{remark}
We compare our scheme to the scheme in \cite{KimLeePSMM}. For any two integers $(m_1,m_2)$, the scheme in \cite{KimLeePSMM} uses $N=(m_1+1)(m_2+1)$ servers to achieve a lower convex hull of $(U,D)=(N/m_1,N/m_1m_2)$. In the simulation, we let $N=12,M=6$ and $K=2,\dots,12$ for our proposed scheme. We used $(m_1,m_2)=(1,5),(2,3),(3,2),(5,1)$ for the scheme in \cite{KimLeePSMM} to ensure that $N=12$. The scheme in \cite{KimLeePSMM} essentially achieves $(U,D)=(N/(K-1),(K/(K-1))(N/(N-K)))$ for $K=2,3,4,6$ in this example. 
In Fig. \ref{fig:compare}, we compare the proposed scheme of this paper with that in \cite{KimLeePSMM}. We next make the following observations: for the same upload cost, the download cost of the proposed scheme is  smaller. 
Furthermore,  there are several $(U,D)$ pairs that the scheme in \cite{KimLeePSMM} cannot achieve, particularly for smaller values of upload cost. 
\end{remark}

\subsection{Illustrative Example: $(N=4,M=2, K=3)$}
We next demonstrate the proposed achievable scheme through an example to illustrate the main ideas. Suppose there are $N=4$ non-colluding servers and each server has access to all $M=2$ matrices, namely $B_1$ and $B_2$, whose dimensions are all $d_2\times d_3$. Assume that the user wants to compute $AB_1$, and the index $\theta=1$ must be kept private. The user first partitions the matrix $A$ into $A=[A_1^T~A_2^T]^T$, where each $A_i$ is of size $(d_1/2)\times d_2,~i=1,2$. The desired computation $AB_1$ becomes $AB_1=[(A_1B_1)^T~(A_2B_1)^T]^T$. To provide security, $A_i$'s are encoded using a secure $(N,K)=(4,3)$ MDS code as follows:
\begin{align}
\widetilde{A}_n=A_1 + A_2x_n + Rx_n^2,~n=1,\dots,4,
\end{align}
where $R\in \mathbb{F}^{(d_1/2)\times d_2}$ is a random matrix (independent of $A$), whose entries are i.i.d. uniform random variables, and $x_n$ is a distinct element in $\mathbb{F}$ assigned to server $n$. The user then sends $\widetilde{A}_n$ to server $n$ and instructs them to multiply their respective $\widetilde{A}_n$ with all $B_m$'s. Specifically, each server $n$ computes
\begin{align}
\widetilde{A}_nB_m &= A_1B_m + A_2B_mx_n + RB_mx_n^2,~m=1,2.
\end{align}
To simplify the notation, we let $h_n^TW^{(m)}\triangleq \widetilde{A}_nB_m$, where
\begin{align}
h_n=\begin{bmatrix}
1\\
x_n\\
x_n^2
\end{bmatrix},~
W^{(m)}=\begin{bmatrix}
A_1B_m\\
A_2B_m\\
RB_m
\end{bmatrix}.
\end{align}
Due to the properties of the $(4,3)$ MDS code, we need at least three different $h_n^TW^{(m)}$ (viewed as three evaluations of a polynomial) to decode the desired result $W^{(1)}$ whose components can then be used to recover $AB_1$ (via polynomial interpolation).

However, only downloading $h_n^TW^{(1)}$ from any three servers will clearly violate the privacy constraint. Thus, to retrieve the answers while preserving privacy, we adopt a similar downloading strategy for the PIR problem with coded databases in \cite{Banawan}. The user asks each server $n$ to divide each $h_n^TW^{(m)}$,$~m=1,2$ into $N^M=4^2=16$ blocks vertically, i.e.,
\begin{align}
h_n^TW^{(m)}=\begin{bmatrix}
h_n^TW_{1}^{(m)}\\
\vdots\\
h_n^TW_{16}^{(m)}
\end{bmatrix},\forall n,m.
\end{align}
Note that each block is composed of rows of the results, each one of which is coded using $(4,3)$ MDS code.

The privacy preserving download scheme is broken into repetitions and rounds. Within each repetition, there are $M=2$ rounds, and there are a total of $K=3$ repetitions. In Repetition $1$, Round $1$, the user downloads $3$ blocks of the desired computation $W^{(1)}$ from each server, i.e., $h_1^TW_{[1:3]}^{(1)}$ from server $1$, $h_2^TW_{[4:6]}^{(1)}$ from server $2$ and so on. To maintain privacy, the user needs to download equivalent amount of blocks of the undesired computation $W^{(2)}$. In order to utilize the undesired blocks, the user needs to be able to decode the undesired blocks. Hence, the user downloads $h_1^TW_{1}^{(2)},h_2^TW_{1}^{(2)}$ and $h_3^TW_{1}^{(2)}$ from server $1,2$ and $3$, respectively; downloads $h_4^TW_{2}^{(2)},h_1^TW_{2}^{(2)}$ and $h_2^TW_{2}^{(2)}$ from server $4,1$ and $2$ respectively, and so on, until all $W_{[1:4]}^{(2)}$ can be decoded (see Table \ref{table:code(4,3)}).

In Round $2$, we pair up a new block of $W^{(1)}$ and a undesired block of $W^{(2)}$ decoded from Round $1$ and let the user download the sum of them. Since the user has not downloaded $W_4^{(2)}$ from server $1$, the user can ask server $1$ to send $h_1^TW_4^{(2)}+h_1^TW_{13}^{(1)}$. Similarly, the user can ask server $2$ to send $h_2^TW_3^{(2)}+h_2^TW_{14}^{(1)}$ and so on. Clearly, the user can use $W_{[1:4]}^{(2)}$ from Round $1$ as side information to obtain $h_1^TW_{13}^{(1)},h_2^TW_{14}^{(1)},h_3^TW_{15}^{(1)}$ and $h_4^TW_{16}^{(1)}$. In Repetition $1$, the user ends up downloading all the blocks of $W^{(1)}$ once ($12$ from Round $1$, and $4$ from Round $2$). In order to decode $W^{(1)}$, the user needs to download all $16$ blocks two more times with different linear combinations. Thus, the user can follow the same downloading pattern with a right shift of indices in Repetition $2$ and one more shift in Repetition $3$ with new side information. The normalized upload and download costs of the scheme are given by $U=4/(3-1)=2$ and $D=(3\times 28)/(2\times 16)=21/8$ per bit. We next describe the general scheme in the following section.

\begin{figure}[t]
	\centering
	\includegraphics[width=0.43\linewidth]{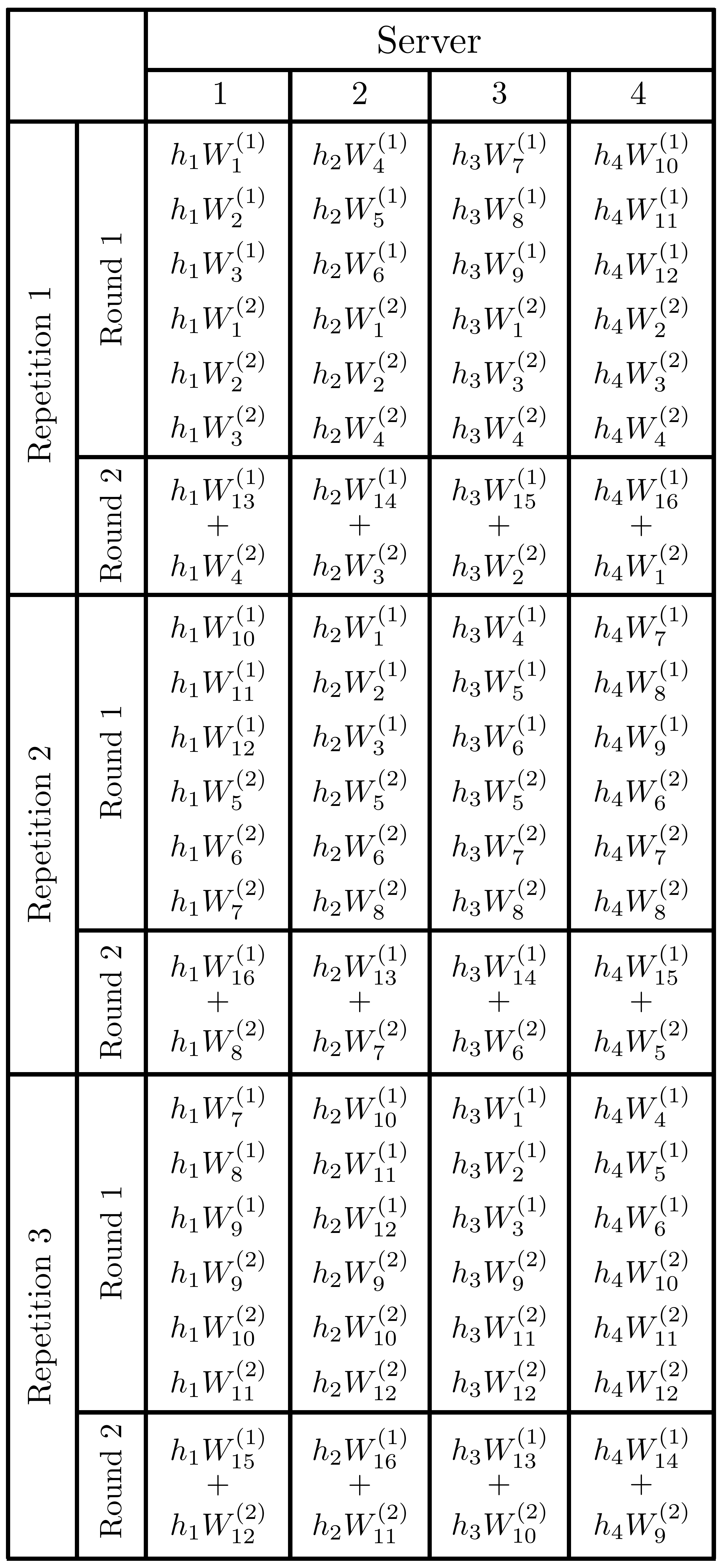}
	\caption{Downloading strategy for $(N=4,M=2, K=3)$. \label{table:code(4,3)}}
	\vspace{-20pt}
\end{figure}

\section{Proof of Theorem \ref{thm:updown}}
We now present the achievable scheme, which can be broken into two phases: $(1)$ secure upload; and $(2)$ private download.

\noindent \textbf{Phase $\mathbf{1}$} \textit{(Secure Upload):} Assume that the user wants to compute $AB_\theta,\theta\in\{1,\dots,M\}$. For a given parameter $K$, the user first divides the input matrix $A$ into $K-1$ partitions vertically into $A=[A_1^T~\dots~A_{K-1}^T]^T$, where each $A_i$ is of size $d_1/(K-1)\times d_2$. The goal can be written as
\begin{align}
AB_\theta = \begin{bmatrix}
A_1B_\theta\\
\vdots\\
A_{K-1}B_\theta
\end{bmatrix}.
\end{align}
The confidential matrix $A$ is encoded as follows:
\begin{align}
\widetilde{A}_n= \sum\limits_{i=1}^{K-1}A_ix_n^{i-1} + Rx_n^{K-1}
\end{align}
for server $n$, where $R$ is a random matrix with the same dimension as any $A_i$, with uniformly distributed i.i.d. entries, and $x_n$ is a distinct element in $\mathbb{F}$ assigned to server $n$. The user uploads $\widetilde{A}_n$ to server $n$ and instructs each server to multiply their respective $\widetilde{A}_n$ with all the $B_m$'s, i.e.,
\begin{align}
\widetilde{A}_nB_m= \sum\limits_{i=1}^{K-1}A_iB_mx_n^{i-1} + RB_mx_n^{K-1},~\forall m,
\end{align}
which can be written as $h_n^TW^{(m)}$, where
\begin{align}
h_n=\begin{bmatrix}
1\\
x_n\\
\vdots\\
x_n^{K-1}
\end{bmatrix},~
W^{(m)}=\begin{bmatrix}
A_1B_m\\
\vdots\\
A_{K-1}B_m\\
RB_m
\end{bmatrix}.
\end{align}
It is clear that  total normalized upload cost is $U=N/(K-1)$.

\noindent \textbf{Phase $\mathbf{2}$} \textit{(Private Download):} We use a similar downloading technique that was proposed in \cite{Banawan}. Thus, before the user downloads anything from servers, the user first asks each server $n$ to partition the results $\widetilde{A}_nB_m$ into blocks as follows
\begin{align}
h_n^TW^{(m)}=\begin{bmatrix}
h_n^TW_{1}^{(m)}\\
\vdots\\
h_n^TW_{N^M}^{(m)}
\end{bmatrix},\forall m,
\end{align}
where each block is of size $d_1/(N^M(K-1))\times d_3$. The user organizes the download into repetitions and rounds. Within each repetition, there are $M$ rounds, and there are a total of $K$ repetitions. Since the download pattern in each repetition is a cyclic shift of the previous repetition, we focus on describing the details of rounds.

In Round $i=1$ of Repetition $1$, the user downloads $K^{M-1}$ desired blocks from each server, where the user downloads $h_n^TW_{(n-1)K^{M-1}+1}^{(\theta)},\dots,h_n^TW_{nK^{M-1}}^{(\theta)}$ from server $n,\forall n$. A total of $NK^{M-1}$ distinct desired blocks are downloaded in this step. However, in order to ensure privacy, the user needs to enforce message symmetry. In other words, the user needs to download equal number of blocks of each undesired computation $h_n^TW^{(m)},~m=1,\dots,M,m\not=\theta$ from server $n$. While undesired, these blocks can be used as side information in the next round. To ensure that the undesired blocks can be used as side information, the user needs to be able to decode those undesired blocks, i.e., obtain $W_j^{(m)}$ for some $j$ and $m\not=\theta$. Since each block is encoded using $(N,K)$ MDS code, the user needs to download each $h_n^TW_j^{(m)}$ from at least $K$ different servers, for a fixed $m\not=\theta$ and $j$. The user can choose to download $h_n^TW^{(m)}_{j_p}$ from server $n=K(p-1)+1,\dots,Kp \mod N$ and $p=1,\dots,NK^{M-2}$ for all $m\not=\theta$. Note that if $n\mod N=0$, we set that $n$ to $N$. Since each $h_n^TW^{(m)}_{j_p}$ is downloaded from $K$ servers, the user is able to decode $W^{(m)}_{j_p}$. Hence, the user can cancel the contribution of any $W^{(m)}_{j_p}$ in the subsequent rounds. In general, in any given Round $i$, the user downloads $h_{n'}^TW^{(m_1)}_{\ell_{1}}+h_{n'}^TW^{(m_2)}_{\ell_{2}}+\dots+h_{n'}^TW^{(m_i)}_{\ell_{i}}$, where $n'=n,\dots,n+K-1$, $m_1,\dots,m_i\in\{1,\dots,M\}\backslash \theta$ and $\ell_1,\dots,\ell_i\in \{1,\dots,N^M\}$. Thus, the amount of side information available at the user is $N{{M-1}\choose i}K^{M-i-1}(N-K)^{i-1}$.

Since for a particular block of side information, there are $N-K$ servers who did not download that block, we can pair that block with a fresh desired block at those $N-K$ servers. Starting from Round $i+1$, the user downloads the sums of $i+1$ blocks from different computations with the help of side information from last round. The user downloads a fresh block of $h_n^TW^{(\theta)}$ and $i$ blocks of side information decoded from last round, i.e., $h_n^TW^{(\theta)}_{r_1}+h_n^TW^{(m_1)}_{\ell_1}+\dots+h_n^TW^{(m_{i})}_{\ell_{i}}$, where $r_1$ is the new block of $W^{(\theta)}$ from server $n$. This allows the user to obtain $N{{M-1}\choose i}K^{M-i-1}(N-K)^{i}$ fresh desired blocks.

The user can repeat this process $K$ times except the indices for desired blocks are shifted right once in every repetition and fresh indices are chosen for side information in every repetition. Since each desired block is downloaded $K$ times from $K$ different source, the user can easily decode $W^{(\theta)}$, hence, recover $AB_\theta$. This allows the user to download a total of $K\sum_{i=1}^{M-1}N{{M-1}\choose i}K^{M-i}(N-K)^{i-1}$ undesired blocks and a total of $K\sum_{i=0}^{M-1}N{{M-1}\choose i}K^{M-i-1}(N-K)^{i}$ desired blocks. Note that each desired block has $K$ terms, however, only $K-1$ of those terms are actually useful due to the addition of the random matrix (for ensuring confidentiality of the matrix $A$). Hence, the download cost for this scheme is
\begin{align}
D&=\frac{\text{total downloaded}}{\frac{K-1}{K}\times\text{total desired}}= \frac{K}{K-1}\left(1 + \frac{\text{total undesired}}{\text{total desired}}\right)\nonumber\\
&= \frac{K}{K-1}\left(1 + \frac{K\sum_{i=1}^{M-1}N{{M-1}\choose i}K^{M-i}(N-K)^{i-1}}{K\sum_{i=0}^{M-1}N{{M-1}\choose i}K^{M-i-1}(N-K)^{i}}\right)\nonumber\\
&= \frac{K}{K-1}\left(1 + \frac{\frac{K}{N-K}\sum_{i=1}^{M-1}{{M-1}\choose i}K^{M-1-i}(N-K)^{i}}{N^{M-1}}\right)\nonumber\\
&= \frac{K}{K-1} \left(1+\frac{\frac{K}{N-K}\left(N^{M-1}-K^{M-1}\right)}{N^{M-1}}\right)\nonumber\\
&= \frac{K}{K-1} \left(1+\frac{K}{N-K}\left(1 - \left(\frac{K}{N}\right)^{M-1}\right)\right)\nonumber\\
&= \frac{K}{K-1} \left(\frac{N-K(\frac{K}{N})^{M-1}}{N-K}\right)= \frac{K}{K-1} \left(\frac{1-(\frac{K}{N})^M}{1-\frac{K}{N}}\right)\nonumber,
\end{align}
which leads to the expression of the normalized download cost $D$ in Theorem \ref{thm:updown}. This completes the proof of Theorem \ref{thm:updown}.

\section{Conclusions \label{Sec:Conclusion}}
In this paper, we studied the problem of secure and private matrix multiplication. We proposed a new scheme that combines ideas from secret sharing and coded PIR. We showed that the lower convex hull of $(U,D)=(N/(K-1),(K/(K-1))(1+(K/N)+\dots+(K/N)^{M-1}))$ for $K=2,\dots,N$ is achievable. We also show that our scheme performs better than the scheme proposed in \cite{KimLeePSMM} in terms of upload and download cost. An interesting future direction is to obtain converse results for the tradeoff between upload/download cost for this problem. 



\bibliographystyle{IEEEtran}
\bibliography{Ref}

\end{document}